\documentclass[amscd,amsmath,amssymb,verbatim, aps, prd, showpacs]{revtex4-2}
\usepackage{amsmath, amsthm, amssymb}

\usepackage{graphicx}
\usepackage{hyperref}
\usepackage{textcomp}
\usepackage[OT1,T1]{fontenc}
\usepackage{epstopdf}

\newtheorem{remark}{Remark}
\newcommand{\wolframcharacter}[1]{}
\newcommand{\beginwidetext}{\begin{widetext}}
\newcommand{\mnras}{Monthly Notices of the Royal Astronomical Society}

\begin{document}

\title{Sensing Gravity with Polarized Electromagnetic Radiation}
\author{Kjell Tangen}
\email{kjell.tangen@gmail.com}
\affiliation{Oslo, Norway}\label{I1}
\date{\today}
\begin{abstract}
Polarization wiggling is an observational effect of a gravitational
field on the polarization of electromagnetic radiation traversing
it. We find that in linear gravity, the polarization wiggle rate
contributions from scalar, vector and tensor perturbations are independent
and gauge invariant. While vector and tensor perturbations do induce
polarization wiggling, scalar perturbations do not. This poses two
natural questions: Can polarized electromagnetic radiation be used
to measure vectorial and tensorial components of gravitational fields
directly? And if so, how? Polarization wiggling is studied for an
arbitrary vector perturbation to the spacetime metric. In a stationary
spacetime, the polarization wiggle rate is proportional to the difference
in frame dragging rate around the direction of propagation between
radiation emission and measurement events. We show how this can
be used to measure the angular momentum of a gravitational source
if the emitter orbits the gravitational source on a known orbit.
Finally, the polarization wiggling effect induced by a gravitational
tensor mode with arbitrary polarization is analyzed. The effect
is demonstrated for two cases: A spacetime with a flat Minkowski
background and an expanding cosmology with a conformally flat background.
In both cases, the polarization wiggling frequency equals the frequency
of the gravitational tensor mode, while the other state parameters
of the gravitational tensor mode are encoded in\ \ the polarization
wiggling amplitude and phase of the polarized radiation. We show
that measurements of polarization wiggling frequency, amplitude
and phase of polarized electromagnetic radiation emitted by multiple
sources at known positions from different directions enables all
state parameters of a gravitational tensor mode to be determined.
\end{abstract}
\maketitle


\section{Introduction}

Electromagnetic radiation is our primary tool for probing the universe,
and its polarization degrees of freedom represent a separate channel
of information about the processes and phenomena with which the
radiation interacts. While gravitational effects on electromagnetic
radiation, like gravitational lensing and redshift, are well understood
and used in astrophysical measurements, gravity's direct effect
on electromagnetic polarization is among the most elusive effects
of gravity. Pending sufficiently accurate techniques for measuring
gravitational effects on polarized radiation, it is important to
understand them and how they can be used as separate means of probing
gravitational fields, possibly beyond what is possible to do by
other means. 

The effect of a gravitomagnetic field on electromagnetic polarization
has been studied extensively from a theoretical perspective by many
authors over many years \cite{Skrotskii-1957,Dehnen-1973,Mashhoon-1973,Perlick-1993,Balazs-1958,Plebanski-1960,Godfrey-1970,Pineault-1977,Su-Mallet-1980,Fayos-Llosa-1982,Ishihara-1988,Gnedin-Dymnikova-1988,Kobzarev-1988,Kopeikin-2001,Nouri-Zonoz-1999,Sereno-2004,Sereno-2005,Brodutch-2011-11,Brodutch-2011-12,Montanari-1998,Ruggiero-2007,Morales-2007}.
The effect is known as the {\itshape Gravitational Faraday Effect},
alternately as the {\itshape Skrotskii Effect} after G. V. Skrotskii,
who first described the effect \cite{Skrotskii-1957}. From an observational
perspective, however, the Gravitational Faraday Effect on a single
ray of radiation provides an insufficient view of the effect of
gravity on electromagnetic polarization, because it lacks unambiguous
quantification \cite{Brodutch-2011-11,Tangen-2024-1}.

{\itshape Polarizaton wiggling}, on the other hand, provides an
unambiguous quantification of the effect of gravity on the state
of polarization of electromagnetic radiation because it is defined
in distinctly operational terms \cite{Tangen-2024-1}. It represents
the effect from the perspective of a single inertial observer: Given
a pointlike inertial emitter of polarized electromagnetic radation,
it uniquely defines how the presence of a gravitational field along
the trajectory of the radiation is manifested in observations made
by this observer. Polarization wiggling manifests as a change in
the measured state of polarization between subsequent observation
events by a single inertial detector. For linearly or elliptically
polarized radiation, it will be seen as a motion of the polarization
axis in the frame of observation. Polarizaton wiggling is by definition
an observable gravitational effect on the state of electromagnetic
polarization. It is induced by gravitational fields present along
the radiation path between the radiation source and the detector.

The polarization wiggling effect was described in a recent paper
for a general 4-dimensional spacetime and electromagnetic radiation
with an arbitrary state of polarization and frequency composition
\cite{Tangen-2024-1}. The present paper follows up by applying these
results to study the effect of weak gravitational fields on electromagnetic
polarization in more detail. 

The main objective of this paper is, by the use of linear gravity
in spacetimes with flat or conformally flat backgrounds, to elucidate
the polarization wiggling effect through a study of the effect of
scalar, vector and tensor metric perturbations on electromagnetic
polarization. We will show that polarization wiggling is not sensitive
to scalar metric perturbations, while vector and tensor perturbations
yield independent, gauge invariant contributions to the polarization
wiggle rates.

The polarization wiggling effect is reviewed in Section \ref{XRef-Section-6121013}.
Section \ref{XRef-Section-4563844} is an overture to the rest of
the paper in which we study how the polarization wiggle rates transform
under arbitrary conformal transformations. It is shown how the\ \ polarization
wiggle rate in the physical frame is related to the polarization
wiggle rate in the presumably much simpler conformal frame. This
greatly simplifies later calculations and enables a study of polarization
wiggling for spacetimes with conformally flat background geometries,
relevant in a cosmological setting.

In Section \ref{XRef-Section-5622558}, we commence our study of
polarization wiggling in linear gravity. In linear gravity, the
gravitational field can be described as a perturbation to a homogeneous
and isotropic spacetime geometry referred to as the {\itshape background
geometry}. When linearizing a metric theory of gravity, the gravitational
field can be decomposed into scalar, vector and tensor perturbations
that are sourced and evolve independently. In Section \ref{XRef-Section-56221022},
we show how this metric decomposition leads to a gauge invariant
decomposition of the polarization wiggle rate into independent components
induced by scalar, vector and tensor perturbations separately. 

Section \ref{XRef-Section-55221021} considers the effect of scalar
perturbations to the spacetime metric. It is shown that scalar perturbations
do not induce polarization wiggling.

In Section \ref{XRef-Section-56221338}, we consider the effect of
general vector perturbations, and show how the polarization wiggle
rate is related to the gravitomagnetic field. As a particular example,
we calculate the effect of a stationary, axi-symmetric gravitational
field induced by a rotating body on polarized electromagnetic radiation
emitted by an emitter orbiting the gravitational source. We show
how a series of polarization wiggle rate measurements of radiation
emitted at different positions along the orbit will enable the angular
momentum of the gravitational source to be determined.

Finally, in Section \ref{XRef-Section-4565521}, we consider the
effect of a gravitational tensor mode with an arbitrary state of
polarization. Two particular examples are shown: the effect of a
gravitational tensor mode propagating in a spacetime with a flat
Minkowski background geometry and a gravitational tensor mode propagating
in an expanding cosmology with a conformally flat background geometry.
We show how radiation from multiple polarized sources can be used
to determine all state parameters of the gravitational tensor mode.

Our metric convention is $(-+++)$, and relativistic units with $c=1$
are used. The Einstein summation convention is assumed on repeated
indices.

\section{Polarization Wiggling}\label{XRef-Section-6121013}

This section briefly reviews the {\itshape polarization wiggling
effect}, described in a recent paper \cite{Tangen-2024-1}. Both
nomenclature and notation introduced there will be used in this
paper. For simplicity, our attention will be restricted to plane
electromagnetic waves with frequencies above the geometric optics
limit. The results generalize to composite electromagnetic radiation
with arbitrary state of polarization and frequency spectrum, given
that the emitted frequency spectrum has negligible power below the
geometric optics limit and that the radiation is emitted with a
constant frequency spectrum \cite{Tangen-2024-1}. 

Polarization wiggling is a motion (wiggling) of the polarization
axis of electromagnetic radiation, as observed by an observer in
a frame of observation that is parallel transported along with the
observer. Polarization wiggling is a pure gravitational effect that
is induced by the gravitational fields present along the null-geodesic
segment between the emitter and the detector.

First, let us assume the presence of a ray of polarized electromagnetic
radiation, represented as a plane electromagnetic wave. In the geometric
optics limit, it can be represented by a vector potential 
\[
A_{\mu }=a_{\mu }e^{i \vartheta },
\]

where $a_{\mu }$ is a slowly varying complex amplitude and $\vartheta
$ is a rapidly varying real phase. Its wave vector $p_{\mu }=\nabla
_{\mu }\vartheta $ is a null vector that satisfies the null geodesic
equation $\nabla _{p}p^{\mu }=0$. The {\itshape magnitude} of the
vector potential is the amplitude $a\equiv {(A_{\mu }^{*}A^{\mu
})}^{1/2}$. The {\itshape polarization vector} $f^{\mu }\equiv a^{\mu
}/a$ is a spacelike, complex unit vector that is parallel transported
along the radiation geodesic.

\,Next, we will assume the presence of a congruence of timelike
curves with unit tangent vectors $u^{\mu }$. They will be referred
to as the {\itshape rest frame congruence}, or {\itshape rest frame},
for short. Based on this timelike congruence, a standard 3+1 decomposition
of the spacetime can be done. With this decomposition, the {\itshape
spatial projector} is defined by
\begin{equation}
\gamma _{\nu }^{\mu }=\delta _{\nu }^{\mu }+u^{\mu }u_{\nu }.%
\label{XRef-Equation-2268398}
\end{equation}

Then, turning our attention to the null geodesic congruence $p^{\mu
}$, at any point along one of its geodesics, the null vector $p^{\mu
}$ can be decomposed into timelike and spacelike vectors as follows:
\begin{equation}
p^{\mu }=p( u^{\mu }+{\hat{p}}^{\mu }) ,%
\label{XRef-Equation-62782635}
\end{equation}

where $p\equiv -u_{\mu }p^{\mu }$ is the photon energy measured
in the rest frame, while $\hat{p}$ is the vector
\begin{equation}
{\hat{p}}^{\mu }\equiv \frac{p^{\mu }}{p}-u^{\mu }=\gamma _{\nu
}^{\mu }\frac{p^{\nu }}{p}.%
\label{XRef-Equation-24172053}
\end{equation}

${\hat{p}}^{\mu }$ is a spacelike unit vector. It is orthogonal
to $u^{\mu }$, and represents the spatial direction of propagation
of the radiation. The scalar momentum can be found by contracting
it with the 4-momentum: $p=p_{\mu }{\hat{p}}^{\mu }$. 
\begin{remark}

Given a null geodesic congruence $p^{\mu }$, a {\itshape polarization
basis} is an orthonormal basis $e_{A}^{\mu }, A=1,2$ spanning the
plane transverse to the direction of propagation ${\hat{p}}^{\mu
}$ and normal to $u^{\mu }$. The polarization basis is assumed to
be parallel transported along $u^{\mu }$ \cite{Tangen-2024-1}. Given
a null geodesic congruence $p^{\mu }$ and a polarization basis $e_{A}^{\mu
}$, the {\itshape frame of observation} is then defined as the tetrad
$\{u^{\mu }, {\hat{p}}^{\mu },e_{A}^{\mu }\}, A=1,2$.
\end{remark}
\begin{remark}

The following remark should be made with respect to the rest frame
congruence: Reference \cite{Tangen-2024-1} makes an a priori assumption
that the rest frame congruence is geodesic. However, this assumption
is not needed to prove any of the results of reference \cite{Tangen-2024-1}.
Hence, its results are valid for any timelike congruence. In this
paper, we will therefore, unless otherwise stated, allow the rest
frame congruence $u^{\mu }$ to be any timelike congruence.\label{XRef-Remark-329163619}
\end{remark}

The {\itshape screen projector }$S_{\mu \nu }$ adapted to the null
geodesic congruence $p^{\mu }$ is defined as
\begin{equation}
S_{\mu \nu }=\gamma _{\mu \nu }-{\hat{p}}_{\mu }{\hat{p}}_{\nu }=g_{\mu
\nu }+u_{\mu }u_{\nu }-{\hat{p}}_{\mu }{\hat{p}}_{\nu },
\end{equation}

while the {\itshape screen rotator} $\epsilon _{\mu \nu }$ is defined
as
\begin{equation}
\epsilon _{\mu \nu }\equiv \epsilon _{\alpha \beta \mu \nu }u^{\alpha
}{\hat{p}}^{\beta }.%
\label{XRef-Equation-427164445}
\end{equation}

$\epsilon _{\alpha \beta \mu \nu }$ is the Levi-Civita tensor. The
screen projector $S_{\mu \nu }$ projects any 4-vector onto the plane
transverse to the direction of propagation $\hat{p}$ (the screen),
while the screen rotator $\epsilon _{\mu \nu }$ is a generator of
rotations in the screen \cite{Tangen-2024-1}.

Expressed in terms of the Riemann tensor $R_{\gamma \rho \beta \sigma
}$, the {\itshape curvature twist }scalar $\mathcal{Z}$ is defined
as
\begin{equation}
\mathcal{Z}\equiv u^{\gamma }{\hat{p}}^{\beta }\epsilon ^{\rho \sigma
}R_{\gamma \rho \beta \sigma }.%
\label{XRef-Equation-3974337}
\end{equation}

The curvature twist $\mathcal{Z}$ can be interpreted as an angular
acceleration of the polarization axis as the radiation propagates
along the geodesic, the {\itshape wiggle acceleration} \cite{Tangen-2024-1}.

By using the relationship between the Riemann and Weyl tensors,
the curvature twist can alternatively be expressed in terms of the
Weyl tensor $C_{\gamma \rho \beta \sigma }$ as
\begin{equation}
\mathcal{Z}=u^{\gamma }{\hat{p}}^{\beta }\epsilon ^{\rho \sigma
}C_{\gamma \rho \beta \sigma }.%
\label{XRef-Equation-55174536}
\end{equation}

The {\itshape transverse polarization vector} $\epsilon ^{\mu }$
is defined as the transverse projection of the polarization vector
$f^{\mu }$: $\epsilon ^{\mu }\equiv S_{\nu }^{\mu }f^{\nu }$. Two
quantities quantifying the gravitational effect on electromagnetic
polarization are the {\itshape polarization axis wiggle rate} $\omega
$ and the {\itshape mean helicity wiggle rate} $\chi $. They are
defined as two transverse projections of the time rate of change
of the transverse polarization vector $\epsilon ^{\mu }$, and therefore
quantify the time rate of change of the observed state of polarization
in a frame of observation \cite{Tangen-2024-1} 
\begin{gather*}
\omega \equiv \mathrm{Re}\left\{  \epsilon ^{\gamma *}\epsilon _{\gamma
\mu }\nabla _{u}\epsilon ^{\mu }\right\} 
\\\chi \equiv -i \epsilon ^{\mu *}\nabla _{u}\epsilon _{\mu }.
\end{gather*}

For linearly polarized radiation or radiation emitted with constant
elliptical polarization, $\omega $ equals the rotation rate of the
polarization axis. $\chi $ quantifies, as the name suggests, the\ \ mean
of the rates of change of the two helicity phases of the radiation.
For radiation emitted with constant elliptical polarization, $\chi
$ is proportional to the rotation rate of the polarization axis.
$\omega $ is independent of both radiation frequency and the emitted
state of polarization, while $\chi $ depends on the degree of circular
polarization, $\mathcal{V}$. $\chi $ vanishes for linearly polarized
radiation, while $\omega $ is undefined for radiation with pure
circular polarization. Their transport equations along the null
geodesic are
\begin{gather}
\nabla _{p}\left( \frac{\omega }{p}\right) =\mathcal{Z}%
\label{XRef-Equation-225164552}
\\\nabla _{p}\left( \frac{\chi }{p}\right) =-\mathcal{V}\mathcal{Z}.%
\label{XRef-Equation-239450}
\end{gather}

The circular polarization degree $\mathcal{V}$ is conserved along
the radiation geodesic \cite{Tangen-2024-1}:
\begin{equation}
\nabla _{p}\mathcal{V}=0.%
\label{XRef-Equation-1012124319}
\end{equation}

By the use of eq. (\ref{XRef-Equation-1012124319}), a linear combination
of eqs. (\ref{XRef-Equation-225164552}) and (\ref{XRef-Equation-239450})
yields
\begin{equation}
\nabla _{p}\left( \frac{\chi +\mathcal{V} \omega }{p}\right) =0.%
\label{XRef-Equation-2423539}
\end{equation}

The general solution to eq. (\ref{XRef-Equation-225164552}) is
\begin{equation}
\omega ( \lambda ) =\frac{p( \lambda ) }{p( \lambda _{*}) }\omega
( \lambda _{*}) +p( \lambda ) \text{}\operatorname*{\int }\limits_{\lambda
_{*}}^{\lambda }d\lambda ^{\prime } \mathcal{Z}( \lambda ^{\prime
}) ,%
\label{XRef-Equation-22775842}
\end{equation}

where $\lambda $ denotes the affine parameter of the null geodesic
of the radiation and $\lambda _{*}$ labels the emission event. Eq.
(\ref{XRef-Equation-2423539}) implies that $\chi ( \lambda ) $ can
be expressed in terms of $\omega ( \lambda ) $:
\begin{equation}
\chi ( \lambda ) =\frac{p( \lambda ) }{p( \lambda _{*}) }\left(
\chi ( \lambda _{*}) +\mathcal{V} \omega ( \lambda _{*}) \right)
-\mathcal{V} \omega ( \lambda ) ,%
\label{XRef-Equation-23113859}
\end{equation}

The $p( \lambda ) /p( \lambda _{*}) $ factor in the first terms
on the right hand side of eqs. (\ref{XRef-Equation-22775842}) and
(\ref{XRef-Equation-23113859}) can be expressed in terms of the
redshift $z( \lambda _{*}) $ of the radiation source:\ \ 
\[
\frac{p( \lambda ) }{p( \lambda _{*}) }={\left( 1+z( \lambda _{*})
\right) }^{-1}.
\]

For radiation emitted with constant polarization, the initial conditions
are $\omega ( \lambda _{*}) =\chi ( \lambda _{*}) =0$. This summarizes
previous results needed for the present paper.

\section{The Scale Dependency of the Polarization Wiggle Rates}\label{XRef-Section-4563844}

Before we address the main topic of this paper, let us start with
a prelude that will be used in later sections. We would like to
understand how polarization wiggling is manifested in expanding
cosmologies. In particular, we would like to understand how the
curvature twist and the polarization wiggle rates vary with the
scale factor. 

In order to separate the scale dependency from the dependencies
on other metric components, we will assume that the spacetime geometry
of the physical spacetime $\tilde{\mathcal{M}}$ is conformally related
to another geometry $\mathcal{M}$, referred to as the {\itshape
conformal frame geometry }or simply the {\itshape conformal frame}.
Their metrics are then related by a conformal factor $\Omega ^{2}$:
\[
\tilde{g}=\Omega ^{2}g.
\]

Here, $\tilde{g}$ denotes the metric of the physical spacetime geometry
$\tilde{\mathcal{M}}$, while $g$ represents the metric of the conformal
frame geometry $\mathcal{M}$. For example, Minkowski space is the
conformal frame of the conformally flat background geometries commonly
used in cosmology.

We can make use of two facts to simplify the computation of polarization
wiggle rates: {\itshape 1) Null geodesics are invariant under conformal
transformations}, and {\itshape 2) Maxwell's equations in vacuum
are conformally invariant}. As will be shown shortly, this enables
us to, with a given\ \ physical frame spacetime geometry, do all
the necessary computations in the conformal frame and transform
the results to the physical frame. Since the conformal frame is
typically a simpler geometry than the physical frame, the ability
to compute wiggle rates in the conformal frame offers a significant
simplification of these computations for spacetimes with non-trivial
conformal factors, such as expanding cosmologies. 

First, we need to determine the conformal weights of the quantities
contributing to the polarization wiggle rate. To do so, we need
the relationship between the Christoffel symbols of the two geometries
\cite{Carrol-2004}:
\begin{equation}
{\tilde{\Gamma }}_{\mu \nu }^{\rho }=\Gamma _{\mu \nu }^{\rho }+C_{\mu
\nu }^{\rho },%
\label{XRef-Equation-59125647}
\end{equation}

where
\begin{equation}
C_{\mu \nu }^{\rho }\equiv \Omega ^{-1}( \delta _{\mu }^{\rho }\nabla
_{\nu }\Omega +\delta _{\nu }^{\rho }\nabla _{\mu }\Omega -g_{\mu
\nu }g^{\rho \lambda }\nabla _{\lambda }\Omega ) .%
\label{XRef-Equation-59125658}
\end{equation}

${\tilde{\Gamma }}_{\mu \nu }^{\rho }$ and\ \ $\Gamma _{\mu \nu
}^{\rho }$ denote the Christoffel symbols of the physical and conformal
frame geometries, respectively.

\subsection{Conformal Properties of Relevant Quantities}

It is known that, in 4 dimensions, Maxwell's theory of electromagnetism
in vacuum is conformally invariant. A consequence of this invariance
is that the vector potential and 4-momentum transform as follows
under a conformal transformation with conformal factor $\Omega ^{2}$:
\begin{gather}
{\tilde{A}}_{\mu }=A_{\mu }, {\tilde{A}}^{\mu }=\Omega ^{-2}A^{\mu
}
\\{\tilde{p}}_{\mu }=p_{\mu }, {\tilde{p}}^{\mu }=\Omega ^{-2}p^{\mu
}.%
\label{XRef-Equation-4302220}
\end{gather}

Then, let us consider the rest frame congruence in the physical
frame, ${\tilde{u}}^{\mu }$. Define the rest frame congruence in
the conformal frame as $u^{\mu }=\Omega  {\tilde{u}}^{\mu }$. $u^{\mu
}$ is evidently a timelike unit vector in the conformal frame.

Next, given any null geodesic ${\tilde{k}}^{\mu }$ in the physical
frame, $k^{\mu }=\Omega ^{2}{\tilde{k}}^{\mu }$ is also null and
satisfies the geodesic equation
\[
\nabla _{k}k^{\mu }=0.
\]

This implies that $k^{\mu }$ is a null geodesic in the conformal
frame. The conformal weights of the quantities that depend on the
two geodesics ${\tilde{p}}^{\mu }$ and ${\tilde{u}}^{\mu }$ can
now be determined:
\begin{gather}
{\tilde{u}}_{\mu }=\Omega  u_{\mu }, {\tilde{u}}^{\mu }=\Omega ^{-1}u^{\mu
}%
\label{XRef-Equation-22765614}
\\\tilde{p}=\Omega ^{-1}p%
\label{XRef-Equation-22774932}
\\{\tilde{\hat{p}}}_{\mu }=\Omega  {\hat{p}}_{\mu }, {\tilde{\hat{p}}}^{\mu
}=\Omega ^{-1}{\hat{p}}^{\mu }%
\label{XRef-Equation-5182829}
\\{\tilde{S}}_{\mu \nu }=\Omega ^{2}S_{\mu \nu }, {\tilde{S}}^{\mu
\nu }=\Omega ^{-2}S^{\mu \nu }
\\{\tilde{\epsilon }}_{\mu \nu }=\Omega ^{2}\epsilon _{\mu \nu },
{\tilde{\epsilon }}^{\mu \nu }=\Omega ^{-2}\epsilon ^{\mu \nu }.
\end{gather}

The transverse polarization vector $\epsilon ^{\mu }$ transforms
as a unit vector:
\[
{\tilde{\epsilon }}_{\mu }=\Omega  \epsilon _{\mu }, {\tilde{\epsilon
}}^{\mu }=\Omega ^{-1}\epsilon ^{\mu }.
\]

The polarization degree $\mathcal{P}$ and the circular polarization
degree $\mathcal{V}$ are invariant under conformal transformations:
\begin{gather}
\tilde{\mathcal{P}}=\mathcal{P}
\\ \tilde{\mathcal{V}}=\mathcal{V}.
\end{gather}

\subsection{Conformal Transformation of the Curvature Twist}

The Weyl tensor transforms as follows under conformal transformations
\cite{Carrol-2004}:
\begin{equation}
{{\tilde{C}}^{\rho }}_{\sigma \mu \nu }={C^{\rho }}_{\sigma \mu
\nu }.%
\label{XRef-Equation-22765628}
\end{equation}

Using eqs. (\ref{XRef-Equation-22765614})-(\ref{XRef-Equation-22765628}),
we obtain the conformal transformation rule of the curvature twist,
as expressed by eq. (\ref{XRef-Equation-55174536}):
\begin{equation}
\tilde{\mathcal{Z}}=\Omega ^{-2}Z.%
\label{XRef-Equation-39172412}
\end{equation}

Eq. (\ref{XRef-Equation-39172412}) is remarkably simple. It complies
with the intuitive notion that, if you scale a geometry by a constant,
coordinate-independent scale factor $\Omega $, the components of
the Riemann tensor will be scaled by a factor $\Omega ^{-2}$.

\subsection{The Scale Dependency of the Polarization Wiggle Rates}

It is now straight forward to determine how the polarization wiggle
rates $\omega $ and $\chi $ behave under conformal transformations.
From eqs. (\ref{XRef-Equation-4302220}) and (\ref{XRef-Equation-39172412})
follows that, if $\omega /p$ and $\chi /p$ are solutions to eqs.
(\ref{XRef-Equation-225164552}) and (\ref{XRef-Equation-239450})
in the conformal frame, so are $\tilde{\omega }/\tilde{p}$ and $\tilde{\chi
}/\tilde{p}$. This implies that, given the same initial conditions,
\begin{gather*}
\frac{\tilde{\omega }}{\tilde{p}}=\frac{\omega }{p}
\\\frac{\tilde{\chi }}{\tilde{p}}=\frac{\chi }{p}.
\end{gather*}

By using eq. (\ref{XRef-Equation-22774932}), we find the polarization
wiggle rates in the physical frame in terms of the conformal frame
values:
\begin{gather}
\tilde{\omega }=\Omega ^{-1}\omega %
\label{XRef-Equation-2484348}
\\\tilde{\chi }=\Omega ^{-1}\chi .%
\label{XRef-Equation-2484412}
\end{gather}

Eqs. (\ref{XRef-Equation-2484348}) and (\ref{XRef-Equation-2484412})
are generally valid for any spacetime with an arbitrary 4-dimensional
conformal frame geometry. It is a remarkably simple result that
enables us to compute the wiggle rates in the presumably much simpler
conformal frame and transform them to the physical frame.

\subsection{How to Apply the Results}

The results derived above are general and apply to any 4-dimensional
spacetime geometry and conformal transformation. Let us lay out
a method to apply the results by taking a look at an example. 

The geometry of an expanding cosmological spacetime is commonly
represented by a metric 
\[
{\tilde{g}}_{\mu \nu }={a( \eta ) }^{2}g_{\mu \nu },
\]

where $\eta $ is conformal time and the scale factor $a( \eta )
$ acts as a time-dependent conformal transformation that relates
the physical spacetime geometry $\tilde{g}$ to a conformal frame
geometry $g$. By expanding the conformal frame geometry as $g_{\mu
\nu }=g_{\mu \nu }^{(0)}+h_{\mu \nu }$ around a chosen background
geometry $g^{(0)}$, gravitational effects can be analyzed by the
use of linear gravity. The background geometry is usually a geometry
that satisfies the cosmological principle of homogeneity and isotropy.
Let us consider a background geometry that is a maximally symmetric
4-geometry with constant curvature. Three geometries cover all such
cases: Minkowski space as a flat geometry with zero curvature and
$R^{4}$ topology, de Sitter space with positive curvature and topology
$R\times S^{3}$ and anti-de Sitter space with negative curvature
and $R\times H^{3}$ topology. $S^{3}$ and $H^{3}$ denote the 3 dimensional
hypersphere and hyperboloid, respectively. All geometries are conformally
flat, with metric
\[
g_{\mu \nu }( K) ={\left( 1+\frac{1}{4}K( r^{2}-\eta ^{2}) \right)
}^{-2}\eta _{\mu \nu },
\]

where $\eta _{\mu \nu }$ is the Minkowski metric, $K$ is a curvature
constant, $r$ is a radial coordinate and $\eta $ is conformal time
\cite{Stephani-2003}. For $K=0$, $g_{\mu \nu }( K) $ is the Minkowski
metric. For $K>0$, $g_{\mu \nu }( K) $ is a metric of de Sitter
space, while for $K<0$, it is a metric of anti-de Sitter space.

While these geometries have different topologies, and spacetimes
using them as conformal backgrounds have widely different properties,
the conformal flatness of all such spacetimes can be used to simplify
the calculation of polarization wiggling. The reason for this is
that the results of eqs. (\ref{XRef-Equation-39172412}),\ \ \ (\ref{XRef-Equation-2484348})
and (\ref{XRef-Equation-2484412}) allow the curvature twist $\mathcal{Z}$
and the polarization wiggle rates to be computed in the Minkowski
frame and transformed to the physical frame using the conformal
transformation
\[
\Omega ( K) =a( \eta ) {\left( 1+\frac{1}{4}K( r^{2}-\eta ^{2})
\right) }^{-1}.
\]

This example demontrates a\ \ more general fact: The conformal and
physical frames may have different topologies, but the results of
this section still apply. Later sections will demonstrate this method
in more detail for a flat ($K=0$) conformal background.

\subsection{Homogeneous Conformal Transformations}\label{XRef-Subsection-4314111}

In the next sections, we will restrict our attention to spacetimes
with {\itshape homogeneous} (time-dependent) {\itshape conformal
factors}. We define a {\itshape homogeneous conformal transformation}
as a conformal transformation with a conformal factor that for a
given 3+1 spacetime decomposition with spatial projector $\gamma
_{\mu }^{\alpha }$ satisfies
\begin{equation}
\gamma _{\mu }^{\alpha }\nabla _{\alpha }\Omega =0.%
\label{XRef-Equation-22691751}
\end{equation}

This means that the conformal factor does not vary in spatial directions,
and therefore only depends on time when expressed in suitable coordinates:
\[
\Omega =a( \eta ) ,
\]

where $\eta $ is the time coordinate in the conformal frame. Thus,
the time-dependent scale factor $a( \eta ) $ used to represent expanding
cosmologies is one example of a homogeneous conformal factor.

In the following, we will refer to a {\itshape homogeneous conformally
flat} geometry as a geometry that is conformally related to Minkowski
space through a homogeneous conformal transformation. This means
that it has a metric that can be written
\[
g_{\mu \nu }={a( \eta ) }^{2}\eta _{\mu \nu }.
\]

In the following sections, the results of this section will be applied
to spacetimes with homogeneous conformally flat backgrounds.

\section{Polarization Wiggling in Linear Gravity}\label{XRef-Section-5622558}

In the following, we will apply linear gravity to analyze the polarization
wiggling effect in spacetimes with homogeneous conformally flat
background geometries, cf. Section \ref{XRef-Subsection-4314111}
above. The physical metric $\tilde{g}$ can be written
\begin{equation}
{\tilde{g}}_{\mu \nu }={a( \eta ) }^{2}g_{\mu \nu }={a( \eta ) }^{2}\left(
\eta _{\mu \nu }+h_{\mu \nu }\right) ,%
\label{XRef-Equation-218101532}
\end{equation}

where $g_{\mu \nu }\equiv \eta _{\mu \nu }+h_{\mu \nu }$ denotes
the conformal frame metric, $\eta _{\mu \nu }$ is the Minkowski
metric, $h_{\mu \nu }$ a metric perturbation in the conformal frame,
$\eta $ denotes conformal time (time in the conformal frame) and
$a( \eta ) $ is a homogeneous (time dependent) scale factor.

By applying the results of the previous section, we find that this
way of relating the physical spacetime to a geometry with a flat
background through a conformal transformation allows us to study
polarization wiggling in the presumably much simpler conformal frame
geometry and use the results of the previous section to transform
the results to the physical frame. This approach offers significant
simplifications to the analysis of polarization wiggling in the
physical frame. For instance, by applying eq. (\ref{XRef-Equation-2484348}),
the polarization axis wiggle rate $\tilde{\omega }$ in the physical
frame can be determined from the quantity in the conformal frame:
\begin{equation}
\tilde{\omega }( \eta ) ={a( \eta ) }^{-1}\omega ( \eta ) .%
\label{XRef-Equation-41415526}
\end{equation}

In linear gravity with a Minkowski background, the linearized Riemann
tensor is \cite{Carrol-2004}
\begin{equation}
R_{\mu \sigma \alpha \lambda }=\frac{1}{2}\left( h_{\mu \lambda
,\sigma \alpha }+h_{\sigma \alpha ,\mu \lambda }-h_{\mu \alpha ,\sigma
\lambda }-h_{\sigma \lambda ,\mu \alpha }\right) +\mathcal{O}( h^{2})
.%
\label{XRef-Equation-24151613}
\end{equation}

The curvature twist derived from eq. (\ref{XRef-Equation-24151613})
is
\begin{equation}
\mathcal{Z}=u^{\left[ \mu \right. }{\hat{p}}^{\left. \alpha \right]
}\epsilon ^{\beta \lambda }h_{\alpha \beta ,\lambda \mu }.%
\label{XRef-Equation-73171623}
\end{equation}

By the use of eq. (\ref{XRef-Equation-24172053}), eq. (\ref{XRef-Equation-73171623})
can be rewritten as follows:
\begin{equation}
\mathcal{Z}=\frac{1}{p}\left( \nabla _{u}\nu ^{\left( p\right) }-\nabla
_{p}\nu ^{\left( u\right) }\right) =\nabla _{u}\nu ^{\left( \hat{p}\right)
}-\nabla _{\hat{p}}\nu ^{\left( u\right) },%
\label{XRef-Equation-73172938}
\end{equation}

where
\begin{equation}
\nu ^{\left( q\right) }\equiv \frac{1}{2}\epsilon ^{\sigma \lambda
}q^{\alpha }h_{\alpha \sigma ,\lambda }%
\label{XRef-Equation-814104233}
\end{equation}

for an arbitrary 4-vector $q^{\mu }$. While the null geodesic is
parametrized by the affine parameter $\lambda $, the timelike rest-frame
geodesics are parametrized by proper time $\tau $. Then, $\nabla
_{p}\nu ^{(u)}=d\nu ^{(u)}/d\lambda $, while $\nabla _{u}\nu ^{(p)}=d\nu
^{(p)}/d\tau $. Using that to lowest order, $d\nu ^{(p)}/d\tau =\partial
\nu ^{(p)}/\partial \eta $, eq. (\ref{XRef-Equation-73172938}) can
be approximated as
\begin{equation}
\mathcal{Z}=\frac{1}{p}\left( \frac{\partial }{\partial \eta }\nu
^{\left( p\right) }-\frac{d\nu ^{\left( u\right) }}{d\lambda }\right)
+\mathcal{O}( h^{2}) .%
\label{XRef-Equation-73173913}
\end{equation}

The second term on the right-hand sidef of eq. (\ref{XRef-Equation-73173913})
can readily be integrated, and using this expression for the curvature
twist in eqs. (\ref{XRef-Equation-22775842}) and (\ref{XRef-Equation-23113859})
gives the following expressions for the wiggle rates:
\begin{gather}
\omega ( \lambda ) =\omega ( \lambda _{*}) +\operatorname*{\int
}\limits_{\lambda _{*}}^{\lambda }d\lambda ^{\prime }\frac{\partial
\nu ^{\left( p\right) }}{\partial \eta }-{\Delta \nu }^{\left( u\right)
}%
\label{XRef-Equation-822175259}
\\\chi ( \lambda ) =-\mathcal{V} \omega ( \lambda ) +\chi ( \lambda
_{*}) +\mathcal{V} \omega ( \lambda _{*}) ,%
\label{XRef-Equation-2423228}
\end{gather}

where
\[
{\Delta \nu }^{\left( u\right) }\equiv \nu ^{\left( u\right) }(
\lambda ) -\nu ^{\left( u\right) }( \lambda _{*}) .
\]

In eqs. (\ref{XRef-Equation-822175259}) and (\ref{XRef-Equation-2423228}),
we used that to lowest order with a Minkowski background, $p( \lambda
) /p( \lambda _{*}) =1 + \mathcal{O}( h) $.

\section{Scalar-Vector-Tensor Decomposition}\label{XRef-Section-56221022}

In linear gravity, the metric perturbation $h_{\mu \nu }$ can be
decomposed into irreducible scalar, vector and tensor components
\cite{lifshitz1946gravitational,Bertschinger:2001is,Poisson-Will-2014}.
This means that, in any metric gauge, we may separate an arbitrary
metric perturbation $h_{\mu \nu }$ into independent scalar, vector
and tensor perturbations:
\begin{equation}
h_{\mu \nu }=h_{\mu \nu }^{\left( S\right) }+h_{\mu \nu }^{\left(
V\right) }+h_{\mu \nu }^{\left( T\right) }.%
\label{XRef-Equation-218101151}
\end{equation}

Since scalar, vector and tensor perturbations are sourced and evolve
independently in linear gravity, their effects can be analyzed separately.
Therefore, since the curvature twist, as expressed by eq. (\ref{XRef-Equation-73171623}),
is linear in the metric perturbation $h_{\mu \nu }$, we may split
both the curvature twist and polarization wiggle rates into scalar,
vector and tensor terms that are induced by corresponding scalar,
vector and tensor perturbations to the metric:
\begin{gather*}
\mathcal{Z}=Z^{\left( S\right) }+Z^{\left( V\right) }+Z^{\left(
T\right) }
\\\omega =\omega ^{\left( S\right) }+\omega ^{\left( V\right) }+\omega
^{\left( T\right) },
\end{gather*}

where
\begin{gather}
\mathcal{Z}^{\left( Q\right) }\equiv u^{\left[ \mu \right. }{\hat{p}}^{\left.
\alpha \right] }\epsilon ^{\beta \lambda }h_{\alpha \beta ,\lambda
\mu }^{\left( Q\right) }%
\label{XRef-Equation-715142842}
\\\omega ^{\left( Q\right) }( \lambda ) =p\text{}\operatorname*{\int
}\limits_{\lambda _{*}}^{\lambda }d\lambda ^{\prime } \mathcal{Z}^{\left(
Q\right) }( \lambda ^{\prime }) %
\label{XRef-Equation-82184114}
\end{gather}

for tensorial order $Q=S,V,T$.

\subsection{Gauge Invariance of Curvature Twist}

The curvature twist $\mathcal{Z}$ is by definition a scalar quantity,
which means it is gauge invariant in linear gravity for an arbitrary
metric perturbation $h_{\mu \nu }$. Since the metric perturbations
$h_{\mu \nu }^{(S)}, h_{\mu \nu }^{(V)}$ and $h_{\mu \nu }^{(T)}$
can be treated as independent perturbations, the corresponding curvature
twist components $\mathcal{Z}^{(S)}, \mathcal{Z}^{(V)}$ and $\mathcal{Z}^{(T)}$
are individially gauge invariant. This implies that the corresponding
polarization wiggle rates $\omega ^{(S)}, \omega ^{(V)}$ and $\omega
^{(T)}$ are also gauge invariant. This is an important finding,
because it means that we have found a way to physically quantify
the effects of scalar, vector and tensor perturbations on electromagnetic
polarization independently. Furthermore, we are free to choose metric
gauge to evaluate the curvature twist and polarization wiggle rate
components.

Next, we will choose a particular metric gauge and evaluate $\mathcal{Z}^{(S)},
\mathcal{Z}^{(V)}$ and $\mathcal{Z}^{(T)}$ individually in the chosen
gauge.

\subsection{Metric Decomposition}

Let us do a scalar-vector-tensor decomposition of the metric perturbation
$h_{\mu \nu }$ in the transverse gauge. In this gauge, the metric
takes the form \cite{Carrol-2004}
\begin{gather*}
h_{00}=h_{00}^{\left( S\right) }=-2\psi 
\\h_{0i}=h_{0i}^{\left( V\right) }=w_{i}
\\h_{\mathrm{ij}}=h_{\mathrm{ij}}^{\left( S\right) }+h_{\mathrm{ij}}^{\left(
T\right) }=2\phi  \delta _{\mathrm{ij}}+h_{\mathrm{ij}}^{\mathrm{TT}},
\end{gather*}

where $\psi $ and $\phi $ are two scalar potentials, $w^{i}$ is
a transverse vector and $h_{\mathrm{ij}}^{\mathrm{TT}}$ is a transverse
and traceless tensor. 

\section{Scalar-induced Polarization Wiggling}\label{XRef-Section-55221021}

In the transverse gauge, the scalar component of the curvature twist
is
\[
\mathcal{Z}^{\left( S\right) }={\hat{p}}^{j}{\hat{p}}^{l}\varepsilon
^{\mathrm{ljk}}{\phi  }_{,\mathrm{k0}}.
\]

Since ${\hat{p}}^{j}{\hat{p}}^{l}\varepsilon ^{\mathrm{ljk}}=0$,
we find that $\mathcal{Z}^{(S)}=0$. Consequently,
\begin{equation}
\omega ^{\left( S\right) }=0.
\end{equation}

Since $\mathcal{Z}^{(S)}$ is gauge invariant, we can conclude that
scalar perturbations to the metric do not contribute to polarization
wiggling.

\section{Vector-induced Polarization Wiggling}\label{XRef-Section-56221338}

\subsection{Polarization Wiggling from the Gravitomagnetic Field}

In the transverse gauge, the vector component of the curvature twist
can be written
\[
\mathcal{Z}^{\left( V\right) }=\frac{1}{2}\hat{\text{\boldmath $p$}}\cdot
\text{\boldmath $\nabla $}\left( \hat{\text{\boldmath $p$}}\cdot
\text{\boldmath $\nabla $}\times \text{\boldmath $w$}\right) 
\]

$\text{\boldmath $\nabla $}\times \text{\boldmath $w$}$ is related
to the gravitomagnetic field $\text{\boldmath $\mathcal{B}$}$ as
$\text{\boldmath $\nabla $}\times \text{\boldmath $w$}=\text{\boldmath
$\mathcal{B}$}$ \cite{Poisson-Will-2014}. The gravitomagnetic field
is a gauge invariant quantity. $\mathcal{Z}^{(V)}$ can be expressed
in terms of the spatial gradient of the gravitomagnetic field strength
in the direction of propagation:
\begin{equation}
\mathcal{Z}^{\left( V\right) }=\frac{1}{2}\hat{\text{\boldmath $p$}}\cdot
\text{\boldmath $\nabla $}\mathcal{B}%
\label{XRef-Equation-82091448}
\end{equation}

where $\mathcal{B}\equiv \hat{\text{\boldmath $p$}}\cdot \text{\boldmath
$\mathcal{B}$}$. We will refer to $\mathcal{B}$ as the {\itshape
longitudinal gravitomagnetic field}. The projection of the gravitomagnetic
field $\text{\boldmath $\mathcal{B}$}$ along the direction of propagation
$\hat{p}$ induces a\ \ rotation of the polarization axis of the
radiation as it propagates along its null geodesic. This effect
of the gravitomagnetic field on polarized electromagnetic radiation
is known as the {\itshape Gravitational Faraday Effect,} due to
its analogy with the Faraday Effect, which is the effect an ambient
magnetic field has on the polarization of passing electromagnetic
radiation. In other words, the Gravitational Faraday Effect can
be viewed as polarization wiggling induced by vector perturbations
through the resulting gravitomagnetic field.

The vector-induced polarization wiggle rate, as given by eq. (\ref{XRef-Equation-82184114}),
can then be expressed in terms of the spatial gradient of the longitudinal
gravitomagnetic field:
\[
\omega ^{\left( V\right) }( \lambda ) \equiv \frac{1}{2}p\text{}\operatorname*{\int
}\limits_{\lambda _{*}}^{\lambda }d\lambda ^{\prime } \hat{\text{\boldmath
$p$}}\cdot \text{\boldmath $\nabla $}\mathcal{B}.
\]

Thus, polarization wiggling can only be induced if the longitudinal
gravitomagnetic field possesses a non-zero spatial gradient along
the direction of propagation of the polarized radiation.

Alternatively, we may use eq. (\ref{XRef-Equation-73173913}). The
terms $\nu ^{(u)}$ and $\nu ^{(p)}$ evaluate to
\begin{gather}
\nu ^{\left( p\right) }=-\frac{p}{2}\hat{\text{\boldmath $p$}}\cdot
\text{\boldmath $\nabla $}\times \text{\boldmath $w$}=-\frac{1}{2}p
\mathcal{B}
\\\nu ^{\left( u\right) }=-\frac{1}{2}\hat{\text{\boldmath $p$}}\cdot
\text{\boldmath $\nabla $}\times \text{\boldmath $w$}=-\frac{1}{2}\mathcal{B}.
\end{gather}

The vector component of the curvature twist then evaluates to
\[
\mathcal{Z}^{\left( V\right) }=-\frac{1}{2}\left( \frac{\partial
\mathcal{B}}{\partial \eta }-\frac{1}{p}\frac{d\mathcal{B}}{d\lambda
}\right) ,
\]

where we used that\ \ $\frac{d\mathcal{B}}{d\tau }=\frac{\partial
\mathcal{B}}{\partial \eta }$ to first order in the metric perturbation.
By eq. (\ref{XRef-Equation-822175259}), the polarization axis wiggle
rate then is
\begin{equation}
\omega ^{\left( V\right) }( \lambda ) =\frac{1}{2}\Delta \mathcal{B}-\frac{1}{2}p\operatorname*{\int
}\limits_{\lambda _{*}}^{\lambda }d\lambda ^{\prime }\frac{\partial
\mathcal{B}}{\partial \eta } ,%
\label{XRef-Equation-8221862}
\end{equation}

where
\[
\Delta \mathcal{B}\equiv \mathcal{B}( \lambda ) -\mathcal{B}( \lambda
_{*}) .
\]

We notice that, even if the gravitomagnetic field is stationary
with $\partial \mathcal{B}/\partial \eta =0$, the vectorial contribution
to the polarization axis wiggle rate, $\omega ^{(V)}$, is still
non-zero if the change in the longitudinal gravitomagnetic field
between the emitter and detector is non-zero. The $\Delta \mathcal{B}$
term on the right-hand side of eq. (\ref{XRef-Equation-8221862})
stems from the term $\nu ^{(u)}$, and can be ascribed to the difference
in inertial frame-dragging rate between the emitter and detector
frames. This can be seen as follows: The angular velocity of the
local inertial frame as measured by a distant observer in a rest
frame is \cite{Misner}
\begin{equation}
{\text{\boldmath $\omega $}}_{f}=-\frac{1}{2}\text{\boldmath $\nabla
$}\times \text{\boldmath $w$}=-\frac{1}{2}\mathcal{B}.%
\label{XRef-Equation-1013103552}
\end{equation}

This\ \ rotation of the local inertial frame is the manifestation
of a gravitational effect known as the {\itshape frame dragging}
effect, or {\itshape dragging of inertial frames}. For an emitter
of radiation with polarization that is constant in the frame of
emission, the polarization axis will rotate with the frame of emission
and opposite of the frame of observation: 
\begin{equation}
\omega ^{\left( V\right) }( \lambda ) =\omega _{f}( \lambda _{*})
+\Delta \omega  -\omega _{f}( \lambda ) ,%
\label{XRef-Equation-101395547}
\end{equation}

where $\omega _{f}\equiv \hat{\text{\boldmath $p$}}\cdot {\text{\boldmath
$\omega $}}_{f}$ is the frame dragging rate around the direction
of propagation and
\[
\Delta \omega  \equiv p\operatorname*{\int }\limits_{\lambda _{*}}^{\lambda
}d\lambda ^{\prime }\frac{\partial {\omega }_{f}( \lambda ^{\prime
}) }{\partial \eta }
\]

 is the net gain in angular rotation rate of an axis subject to
an angular acceleration $\partial \omega _{f}/\partial \eta $ along
the radiation trajectory. It is evident from eq. (\ref{XRef-Equation-1013103552})
that $\Delta \omega $ is non-zero only for non-stationary gravitomagnetic
fields. The natural interpretation of eq. (\ref{XRef-Equation-101395547})
is that $\omega ^{(V)}$ is the net angular rotation rate of the
polarization axis after subtracting the angular frame dragging rate
of the frame of observation. 

Let us now consider eq. (\ref{XRef-Equation-101395547}) for a few
special cases. First of all, an interesting consequence of the result
of eq. (\ref{XRef-Equation-101395547}) is that, given a distant
observer at rest observing an inertial emitter of polarized electromagnetic
radiation with constant polarization orbiting a gravitational source
with a stationary gravitomagnetic field, the polarization axis wiggle
rate of the observed polarized radiation is a direct measure of
the angular frame dragging rate of the frame of emission:
\begin{equation}
\omega ^{\left( V\right) }( \lambda ) =\omega _{f}( \lambda _{*})
,%
\label{XRef-Equation-55223342}
\end{equation}

Conversely, given a distant emitter of polarized electromagnetic
radiation with constant polarization at a position with no ambient
gravitomagnetic field and a detector moving inertially in the vicinity
of a gravitational source with a stationary gravitomagnetic field,
the polarization axis wiggle rate of the measured polarized radiation
is a direct measure of the angular frame dragging rate of the frame
of observation:
\begin{equation}
\omega ^{\left( V\right) }( \lambda ) =-\omega _{f}( \lambda ) .
\end{equation}

Finally, consider\ \ eq. (\ref{XRef-Equation-101395547}) for the
case $\omega _{f}( \lambda _{*}) =\omega _{f}( \lambda ) =0$, which
occurs when the gravitomagnetic field is negligible at both the
detector and emitter locations. In this case, $\omega ^{(V)}$ is
non-zero only if the radiation traverses the vicinity of a gravitational
source with a non-stationary gravitomagnetic field. The gravitomagnetic
field is sourced by the matter current $j^{i}\equiv T^{0i}\simeq
\rho  v^{i}$ \cite{Mashhoon:2003ax}, where $T^{\mu \nu }$ is the
energy-momentum tensor, $\rho $ is matter density and $v^{i}$ is
matter velocity. Therefore, changing matter currents induce changes
to the gravitomagnetic field, which in turn induce polarization
wiggling in any passing electromagnetic radiation.

We conclude that the polarization axis wiggle rate $\omega ^{(V)}$
is a direct measure of the longitudinal gravitomagnetic field $\mathcal{B}$.
The contribution to the polarization axis wiggle rate from vector
perturbations has two terms. The first contribution is due to the
difference in the angular frame dragging rate between the frame
of emission and the frame of observation. The second contribution
stems from any non-stationary gravitomagnetic fields present along
the radiation trajectory. 

It is worth remarking that the polarization wiggle rate induced
by a localized gravitomagnetic field on a Minkowski backgrond is
to a large extent insensitive to distance between the emitter and
detector. This can be seen from eq. (\ref{XRef-Equation-8221862}).
The first term on the right-hand side is clearly independent of
distance, while the second term is independent of distance because
$\mathcal{B}$ is non-zero in just a limited part of the path between
emitter and detector.

\subsection{Measuring Angular Momentum of a Gravitational Source
using Polarization Wiggling}

We will now apply the results for arbitrary scalar and vector perturbations
to analyze a more specific case. We will show that the polarization
wiggling of electromagnetic radiation emitted by a radiation emitter
in orbit around a gravitational source with unknown angular momentum
can be used to determine the angular momentum of the gravitational
source.

Let us begin by stating the spacetime metric in the vicinity of
the gravitational source. We will assume it to be a stationary axi-symmetric
metric with a line element that in isotropic coordinates can be
written as \cite{Poisson-2004}
\begin{equation}
{\mathrm{ds}}^{2}=-\left( 1-\frac{2m}{r}\right) {\mathrm{dt}}^{2}+\left(
1+\frac{2m}{r}\right) \left( {\mathrm{dr}}^{2}+r^{2}{\mathrm{d\Omega
}}^{2}\right) -\frac{4j {\sin }^{2}\theta }{r}\mathrm{dt} \mathrm{d\phi
},%
\label{XRef-Equation-43153557}
\end{equation}

where $t$ is time, $(r,\theta ,\phi )$ are spherical coordinates,
$m$ is the mass of the gravitational source and $j$ is the scalar
angular momentum, assuming the rotation axis is the z-axis $\hat{z}$.
Eq. (\ref{XRef-Equation-43153557}) can be found as eq. (4.83) of
ref. \cite{Poisson-2004}. In Section \ref{XRef-Section-55221021},
we found that scalar perturbations do not contribute to polarization
wiggling. The only component of the above metric that contributes
is the $\mathrm{dt} \mathrm{d\phi }$ term. The vector perturbation
is
\[
w_{r}=w_{\theta }=0, w_{\phi }=-\frac{4j {\sin }^{2}\theta }{r}.
\]

The gravitomagentic field $\text{\boldmath $\mathcal{B}$}$ is then
\begin{equation}
\text{\boldmath $\mathcal{B}$}\equiv \text{\boldmath $\nabla $}\times
\text{\boldmath $w$}=-\frac{12j}{r^{2}}\mu \sqrt{1-\mu ^{2}} \text{\boldmath
$n$},%
\label{XRef-Equation-5522187}
\end{equation}

where $\text{\boldmath $n$}$ is the radial unit vector, $\mu \equiv
\hat{\text{\boldmath $j$}}\cdot \text{\boldmath $n$}$, where $\hat{\text{\boldmath
$j$}}=\text{\boldmath $J$}/j$ is a unit vector along the rotation
axis of the gravitational source and $\text{\boldmath $J$}$ is the
angular momentum of the gravitational source. Eq. (\ref{XRef-Equation-5522187})
is a dipole field that vanishes in the equatorial plane. The polarization
wiggle rate for this case is given by eq. (\ref{XRef-Equation-55223342}).
The result is
\begin{equation}
\omega ^{\left( V\right) }( t) =-\frac{1}{2}\hat{\text{\boldmath
$p$}}\cdot \text{\boldmath $\mathcal{B}$}( t_{*}) =\frac{6j}{{r(
t_{*}) }^{2}}\mu ( t_{*}) \sqrt{1-{\mu ( t_{*}) }^{2}} \hat{\text{\boldmath
$p$}}\cdot \text{\boldmath $n$}( {t}_{*}) ,%
\label{XRef-Equation-5662750}
\end{equation}

where $t_{*}$ is time of emission, while $\text{\boldmath $r$}(
{t}_{*}) \text{}=r( t_{*})  \text{\boldmath $n$}( t_{*}) $ is the
orbit of the emitter and $\mu ( {t}_{*}) \equiv \hat{\text{\boldmath
$j$}}\cdot \text{\boldmath $n$}( {t}_{*}) $. We will assume that
the orbit $\text{\boldmath $r$}( {t}_{*}) $ of the emitter is known,
while the angular momentum $\text{\boldmath $J$}$ of the gravitational
source is unknown. There are three unknowns that can be determined
by conducting three measurements of the polarization wiggle rate
$\omega ^{(V)}$ at different points along the orbit of the emitter.
Then eq. (\ref{XRef-Equation-5662750}) can be used to solve for
the three components of the angular momentum $\text{\boldmath $J$}$
of the gravitational source.

\section{Tensor-induced Polarization Wiggling}\label{XRef-Section-4565521}

The tensor perturbation $h_{\mu \nu }^{(T)}$ of eq. (\ref{XRef-Equation-218101151})
is a gauge invariant quantity that is transverse and traceless,
conventionally denoted $h_{\mu \nu }^{\mathrm{TT}}$. To give the
results of the previous sections a proper demonstration, we will
now explore the polarization wiggle rate $\tilde{\omega }$ arising
from a single gravitational tensor mode in linear gravity, assuming
a spacetime with a conformally flat background. This will cover
spacetimes with both a flat Minkowski background as well as expanding
cosmologies.

Let us assume a cosmology with a homogeneous conformally flat background
and the metric of eq. (\ref{XRef-Equation-218101532}) with a transverse,
traceless metric perturbation $h_{\mu \nu }^{\mathrm{TT}}$. The
spatial Einstein equation in vacuum yields in this case \cite{Maggiore-2-Cosmology}
\begin{equation}
{\left( h_{\mu \nu }^{\mathrm{TT}}\right) }^{\prime \prime }+2{\mathcal{H}(
h_{\mu \nu }^{\mathrm{TT}}) }^{\prime }-\nabla ^{2}h_{\mu \nu }^{\mathrm{TT}}=0,%
\label{XRef-Equation-3218016}
\end{equation}

where $\mathcal{H}\equiv a^{\prime }/a$ is the conformal Hubble
factor and ` denotes differentiation with respect to conformal time
$\eta $. We will not attempt to study the effect of a general tensor
perturbation $h_{\mu \nu }^{\mathrm{TT}}$. Instead, under the presumption
that an arbitrary tensor perturbation $h_{\mu \nu }^{\mathrm{TT}}$
can be represented as a linear combination of simple plane wave
components, we will study the effect of a single plane gravitational
wave (tensor mode) with arbitrary polarization. The polarization
wiggle rate indued by a composite gravitational tensor perturbation
can then simply be aggregated as the sum of the polarization wiggle
rates of the plane wave components of the tensor perturbation.

\subsection{Wiggle Rate from a Tensor Mode with a Conformally Flat
Background }

We will now explore the polarization wiggle rate $\tilde{\omega
}$ in the physical frame arising from a single gravitational tensor
mode propagating in an expanding universe with a conformally flat
background geometry. Let us apply the following trial solution to
eq. (\ref{XRef-Equation-3218016}):
\begin{equation}
h_{\mu \nu }^{\mathrm{TT}}=C_{\mu \nu }( \eta ) e^{i k {\hat{k}}_{\mu
}x^{\mu }},%
\label{XRef-Equation-3218520}
\end{equation}

where $\hat{k}$ is a spatial unit vector, and $k>0$ is an arbitrary
angular frequency. Notice that we use a complex representation of
the tensor mode for the sake of analytic and notational simplicity,
with the implicit understanding that only the real part of expressions
derived linearly from it should be used in physical quantities.
${\hat{k}}^{\mu }$ is the direction of propagation, and $k$ is the
angular frequency of the gravitational mode. $C_{\mu \nu }( \eta
) $ is a time-dependent polarization tensor.\ \ 

Eq. (\ref{XRef-Equation-3218520}) is a solution to eq. (\ref{XRef-Equation-3218016})\ \ if
$C_{\mu \nu }$ satisfies
\begin{equation}
C_{\mu \nu }^{\prime \prime }+2\mathcal{H} C_{\mu \nu }^{\prime
}+k^{2}C_{\mu \nu }=0.%
\label{XRef-Equation-41141910}
\end{equation}

We will seek solutions $C_{\mu \nu }( \eta ) \propto e^{-i k \eta
}$. In that case, the metric perturbation of eq. (\ref{XRef-Equation-3218520})
represents a single tensor mode with a null wave vector $k^{\mu
}=k( u^{\mu }+{\hat{k}}^{\mu }) $ and a time-dependent polarization
tensor $C_{\mu \nu }( \eta ) $, where $\hat{k}$ is the direction
of propagation and a spatial unit vector, while $k=-u_{\mu }k^{\mu
}$ is the angular frequency of the tensor mode. The polarization
tensor C${}_{\mu \nu }( \eta ) $ can be expressed in terms of a
tensor basis $\sigma _{\mu \nu }^{s}( \hat{k}) $ for a given direction
of propagation ${\hat{k}}^{\mu }$ as
\[
C_{\mu \nu }( \eta ) =\sum \limits_{s=+,\times }\sigma _{\mu \nu
}^{s}( \hat{k}) c_{s}( k,\eta )  ,
\]

where the sum is over polarizations $s=+,\times $. $c_{s}( k,\eta
) $ is the amplitude of the polarization $s$ component of the tensor
mode. The basis tensors $\sigma _{\mu \nu }^{s}( \hat{k}) $ are
assumed traceless, and they must satisfy $u^{\mu }\sigma _{\mu \nu
}^{s}={\hat{k}}^{\mu }\sigma _{\mu \nu }^{s}=0$ and the normalization
condition $\sigma _{\mu \nu }^{s}( \hat{k}) \sigma ^{s^{\prime }\mu
\nu }( \hat{k}) =2\delta ^{s s^{\prime }}$ \cite{Maggiore-2-Cosmology}.
This ensures that the metric perturbation of eq. (\ref{XRef-Equation-3218520})
is transverse and traceless by construction. 

From eq. (\ref{XRef-Equation-41141910}) follows that the tensor
mode amplitudes must satisfy the differential equation
\begin{equation}
c_{s}^{\prime \prime }+2\mathcal{H} c_{s}^{\prime }+k^{2}c_{s}=0.%
\label{XRef-Equation-32220744}
\end{equation}

First, let us try to determine the amplitude $c_{s}( \eta ) $ by
solving eq. (\ref{XRef-Equation-32220744}) with the following trial
solution:
\begin{equation}
c_{s}( k,\eta ) =h_{s}( k,\eta ) e^{-i k \eta }.%
\label{XRef-Equation-59141220}
\end{equation}

Eq. (\ref{XRef-Equation-32220744}) then takes the form
\begin{equation}
{\left( h_{s}^{\prime }-2i k h_{s}\right) }^{\prime }+2\mathcal{H}(
h_{s}^{\prime }-i k h_{s}) =0.%
\label{XRef-Equation-2168524}
\end{equation}

Let us assume that $h_{s}( \eta ) $ is varying slowly with time,
with $|h_{s}^{\prime }|\ll |k h_{s}|$. In this limit, eq. (\ref{XRef-Equation-2168524})
has the approximate solution
\begin{equation}
h_{s}( k,\eta ) \simeq \frac{h_{s}^{0}( k) }{a( \eta ) },%
\label{XRef-Equation-5914141}
\end{equation}

where $h_{s}^{0}$ is a constant of integration. We have now established
the time dependency of the polarization amplitude $c_{s}( k,\eta
) $ for a tensor mode. Next, let us evaluate the polarization wiggle
rate $\omega $ in the conformal frame.

From eq. (\ref{XRef-Equation-82184114}) follows that the polarization
wiggle rate is
\begin{equation}
\omega ^{\left( T\right) }( \lambda ) \equiv p\text{}\operatorname*{\int
}\limits_{\lambda _{*}}^{\lambda }d\lambda ^{\prime } \mathcal{Z}^{\left(
T\right) }( \lambda ^{\prime }) .%
\label{XRef-Equation-414162935}
\end{equation}

For background null geodesics in the conformal frame, we have $d\eta
/d\lambda =p=\mathrm{const}$. We may use this to reparametrize\ \ both
the radiation and the gravitational mode geodesics in terms of conformal
time. Eq. (\ref{XRef-Equation-414162935}) can then be written 
\begin{equation}
\omega ^{\left( T\right) }( \eta ) =\operatorname*{\int }\limits_{\eta
_{*}}^{\eta }d\eta ^{\prime } \mathcal{Z}^{\left( T\right) }( \eta
^{\prime }) ,%
\label{XRef-Equation-414164019}
\end{equation}

where $\eta _{*}$ is the time of emission. 

We will now evaluate the wiggle rate $\omega $ in the conformal
frame by using eq. (\ref{XRef-Equation-414164019}). The wiggle rate
$\tilde{\omega }$ in the physical frame can then be derived by using
eq. (\ref{XRef-Equation-41415526}). Since we are evaluating $\mathcal{Z}^{(T)}$
to first order in the metric perturbation, we may use the geodesic
in the Minkowski background when evaluating it. Assume that the
observer is at the origin with a conformal frame rest frame geodesic
$x_{0}^{\mu }( \eta ) =\eta  u^{\mu }$. Let the radiation geodesic
in the conformal frame be $x^{\mu }( \lambda ) =x_{0}^{\mu }( \eta
) +p^{\mu }( \lambda -\lambda _{0}) $, where $\lambda _{0}$ labels
the observation event. Then
\[
{\hat{k}}_{\mu }x^{\mu }=-p \mu ( \lambda _{0}-\lambda ) =-\mu (
\eta -\eta ^{\prime }) ,
\]

where $\mu \equiv {\hat{k}}_{\mu }{\hat{p}}^{\mu }$, $\eta $ is
the time of observation and $\eta ^{\prime }$ is the time parameter
of an arbitrary point along the radiation geodesic. $\mu =\cos 
\theta $, where $\theta $ is the angle of inclination of the observed
radiation with the propagation direction of the tensor mode.

Using eq. (\ref{XRef-Equation-715142842}), the curvature twist in
the conformal frame of this metric is
\begin{equation}
\mathcal{Z}^{\left( T\right) }( \eta ^{\prime }) =i k {\hat{p}}^{\mu
}\epsilon ^{\nu \lambda } {\hat{k}}_{\lambda }C_{\mu \nu }^{\prime
}e^{i k {\hat{k}}_{\beta }x^{\beta }}=i k\sum \limits_{s} \sigma
^{s}( \hat{k},\hat{p}) c_{s}^{\prime }e^{-i k \mu ( \eta -\eta ^{\prime
}) },
\end{equation}

where we have introduced the angular function
\begin{equation}
\sigma ^{s}( \hat{k},\hat{p}) \equiv {\hat{p}}^{\mu }\sigma _{\mu
\nu }^{s}( \hat{k})  l^{\nu },%
\label{XRef-Equation-419182926}
\end{equation}

with $l^{\nu }\equiv \epsilon ^{\nu \lambda } {\hat{k}}_{\lambda
}$. We notice that the vector $l^{\nu }$ is transverse to both $\hat{p}$
and $\hat{k}$. The scalar $\sigma ^{s}( \hat{k},\hat{p}) $ represents
the coupling between the polarization $s$ component of a gravitational
tensor mode propagating in direction $\hat{k}$ and the polarization
of electromagnetic radiation propagating in direction $\hat{p}$.
We will refer to $\sigma ^{s}( \hat{k},\hat{p}) $ as the {\itshape
gravitational polarization sensitivity} of polarized radiation emitted
in direction $\hat{p}$ to a gravitational tensor mode with polarization
$s$ propagating in direction $\hat{k}$. It is a purely geometric
quanty that will be analyzed in detail in the next section. The
right-hand side of eq. (\ref{XRef-Equation-414164019}) can be integrated
by parts, and we obtain 
\begin{multline}
\omega ^{\left( T\right) }( \eta ) =\mathrm{Re}\left\{ i k \sum
\limits_{s}\sigma ^{s}( \hat{k},\hat{p}) \left( c_{s}( k,\eta )
-c_{s}( k,\eta _{*}) e^{-i k \mu ( \eta -\eta _{*}) }\right. \right.
\\
\left. \left. -i k \mu  e^{-i k \mu  \eta }\operatorname*{\int }\limits_{\eta
_{*}}^{\eta }d\eta ^{\prime }e^{i k \mu  \eta ^{\prime }}c_{s}(
k,\eta ^{\prime }) \right) \right\} .%
\label{XRef-Equation-31984610}
\end{multline}

Inserting the $c_{s}$ solution of eqs. (\ref{XRef-Equation-59141220})
and (\ref{XRef-Equation-5914141}) into eq. (\ref{XRef-Equation-31984610})
yields the following expression for the polarization wiggle rate:
\begin{multline}
\omega ^{\left( T\right) }( \eta ) =\mathrm{Re}\left\{  i k \sum
\limits_{s}\sigma ^{s}( \hat{k},\hat{p}) h_{s}^{0}( k)  \left( \frac{e^{-i
k \eta }}{a( \eta ) }\right. \right. \\
\left. \left. -e^{-i k \mu  \eta }\left( \frac{e^{-i k( 1-\mu )
\eta _{*}}}{a( \eta _{*}) }+i k \mu  \operatorname*{\int }\limits_{\eta
_{*}}^{\eta }d\eta ^{\prime }\frac{e^{-i k \left( 1-\mu \right)
\eta ^{\prime }}}{a( \eta ^{\prime }) }\right) \right) \right\}
.%
\label{XRef-Equation-4157300}
\end{multline}

Assuming that both the detector and emitter follow rest frame geodesics,
we may assume that the comoving distance between them is constant
to lowest order:
\[
\chi _{*}\equiv \eta -\eta _{*}=\mathrm{const}.
\]

The wiggle rate can then be expressed in terms of comoving distance:
\begin{equation}
\omega ^{\left( T\right) }( \eta ) =\mathrm{Re}\left\{ i k e^{-i
k \eta }\left( \frac{1}{a( \eta ) }-\frac{e^{i k( 1- \mu  ) \chi
_{*}}}{a( \eta _{*}) }-i k \mu \ \ \operatorname*{\int }\limits_{0}^{\chi
_{*}}d\chi \frac{e^{i k \left( 1-\mu \right)  \chi }}{a( \eta -\chi
) }\right)  \sum \limits_{s}\sigma ^{s}( \hat{k},\hat{p}) h_{s}^{0}(
k) \right\} .%
\label{XRef-Equation-217162013}
\end{equation}

Eq. (\ref{XRef-Equation-4157300}) gives the polarization wiggle
rate $\omega $ in the conformal frame for any cosmology with a conformally
flat background. Furthermore, since at present time $\eta =\eta
_{0}$, $a( \eta _{0}) =1$, the present time polarization wiggle
rate in the physical frame, which is given by eq. (\ref{XRef-Equation-41415526}),
is equal to the conformal frame value:
\[
{\tilde{\omega }}^{\left( T\right) }( \eta _{0}) =\omega ^{\left(
T\right) }( \eta _{0}) .
\]

It is apparent from eq. (\ref{XRef-Equation-4157300}) that, if $a(
\eta _{*}) \ll 1$, the second term on the right-hand side will be
the dominant one of the two first terms. This is the case if the
polarized radiation and the tensor mode were emitted at early times
in an expanding cosmology. In the next sections, we will evaluate
eq. (\ref{XRef-Equation-4157300}) for two different backgrounds:
A flat Minkowski background and an expanding, conformally flat background.

\subsection{Gravitational Polarization Sensitivity $\sigma ^{s}$}

Next, let us evaluate the gravitational polarization sensitivity
$\sigma ^{s}( \hat{k},\hat{p}) $ of eq. (\ref{XRef-Equation-419182926}).
It is defined as a purely geometric quantity in terms of two spatial
directions $\hat{k}$ and $\hat{p}$ and the tensor polarization basis
$\sigma _{\mu \nu }^{s}$. First, the tensor basis $\sigma _{\mu
\nu }^{s}( \hat{k}) $ for a given direction of propagation ${\hat{k}}^{\mu
}$ must be chosen. With the choice of an orthonormal polarization
basis $e_{A}^{\mu }( \hat{k}) , A=1,2$ perpendicular to the propagation
direction $\hat{k}$, the basis tensor $\sigma _{\mu \nu }^{s}$ can
be defined by
\[
\sigma _{\mu \nu }^{s}( \hat{k}) \equiv \sigma _{\mathrm{AB}}^{s}e_{\mu
}^{A}( \hat{k}) e_{\nu }^{B}( \hat{k}) ,
\]

where $\sigma _{\mathrm{AB}}^{s}$ are the two symmetric Pauli matrices
\[
\sigma _{\mathrm{AB}}^{+}\equiv \left( \begin{array}{cc}
 1 & 0 \\
 0 & -1
\end{array}\right) , \sigma _{\mathrm{AB}}^{\times }\equiv \left(
\begin{array}{cc}
 0 & 1 \\
 1 & 0
\end{array}\right) .
\]

Then,
\begin{gather}
\sigma _{\mu \nu }^{+}=e_{\mu }^{1}e_{\nu }^{1}-e_{\mu }^{2}e_{\nu
}^{2}%
\label{XRef-Equation-42716501}
\\\sigma _{\mu \nu }^{\times }=e_{\mu }^{1}e_{\nu }^{2}+e_{\mu }^{2}e_{\nu
}^{1}.%
\label{XRef-Equation-427165219}
\end{gather}

Using the definition of the screen rotator in eq. (\ref{XRef-Equation-427164445}),\ \ $\sigma
^{s}$, as defined in eq. (\ref{XRef-Equation-419182926}), can be
expressed as
\begin{gather*}
\sigma ^{+}( \hat{k},\hat{p}) =-{\hat{p}}^{\mu }\sigma _{\mu \nu
}^{\times }{\hat{p}}^{\nu }
\\\sigma ^{\times }( \hat{k},\hat{p}) ={\hat{p}}^{\mu }\sigma _{\mu
\nu }^{+}{\hat{p}}^{\nu }.
\end{gather*}

Furthermore, using eqs. (\ref{XRef-Equation-42716501}) and (\ref{XRef-Equation-427165219}),
we get
\begin{gather}
\sigma ^{+}( \hat{k},\hat{p}) =-2\beta _{1}\beta _{2}%
\label{XRef-Equation-42812371}
\\\sigma ^{\times }( \hat{k},\hat{p}) =\beta _{2}^{2}-\beta _{1}^{2}%
\label{XRef-Equation-428123954}
\end{gather}

where $\beta _{A}( \hat{k},\hat{p}) \equiv {\hat{p}}_{\mu }e_{A}^{\mu
}( \hat{k}) , A=1,2$ are the projections of the radiation propagation
direction $\hat{p}$ against the polarization basis vector $e_{A}(
\hat{k}) $ for direction $\hat{k}$.

If we choose spherical coordinates such that the tensor mode propagates
in the negative z-direction with $\hat{\text{\boldmath $k$}}=(0,0,-1)$,
the polarized electromagnetic radiation propagates in an arbitrary
direction $\hat{p}=-(\cos  \phi  \sin  \theta , \sin  \phi  \sin
\theta , \cos  \phi )$, the gravitational polarization sensitivities
are
\begin{gather*}
\sigma ^{+}( \hat{k},\hat{p}) =-\sin  2\phi  {\sin }^{2}\theta =-\left(
1-\mu ^{2}\right) \sin  2\phi 
\\\sigma ^{\times }( \hat{k},\hat{p}) =\cos  2\phi  {\sin }^{2}\theta
=\left( 1-\mu ^{2}\right) \cos  2\phi .
\end{gather*}

\subsection{Wiggle Rate from a Tensor Mode - Close Emitter}

Let us consider the simplest case first: An emitter of polarized
electromagnetic radiation in the cosmological neighborhood, which
will be referred to as a Close Emitter (CE). In this case, we can
assume a flat background with $a( \eta ) =\mathrm{const} = 1$. When
evaluating the wiggle rate of eq. (\ref{XRef-Equation-4157300})
for this case, the result is
\begin{equation}
\omega _{\left( \operatorname{CE}\right) }^{\left( T\right) }( \eta
) =\mathrm{Re}\left\{ i k\frac{\left( 1-e^{i \left( 1-\mu \right)
k \chi _{*}}\right) }{1-\mu }e^{-i k \eta }\sum \limits_{s} \sigma
^{s}( \hat{k},\hat{p}) h_{s}^{0}( k) \right\} .%
\label{XRef-Equation-62992410}
\end{equation}

To lowest order, proper time of the observer equals coordinate time;
$\tau =\eta +\mathcal{O}( h) $. Since $\omega _{\mathrm{CE}}^{(T)}$
is $\mathcal{O}( h) $, we may interchange proper time and coordinate
time in the above expression. Then, integrating with respect to
proper time $\tau $ gives the polarization angle
\begin{equation}
\Omega _{\left( \operatorname{CE}\right) }^{\left( T\right) }( \tau
) =\mathrm{Re}\left\{ \left( \sigma ^{+}( \hat{k},\hat{p}) h_{+}^{0}(
k) +\sigma ^{\times }( \hat{k},\hat{p}) h_{\times }^{0}( k) \right)
\frac{\left( e^{i k( 1-\mu ) \chi _{*}}-1\right) }{\left( 1-\mu
\right) }e^{-i k \tau }\right\}  +\overline{\Omega },%
\label{XRef-Equation-427221941}
\end{equation}

where the integration constant $\overline{\Omega }$ is the mean
angle, which in the following will be set to zero.

In order to calculate $\Omega _{(\mathrm{CE})}^{(T)}$, we need to
separate the amplitude and phase of each complex factor on the right-hand
side of eq. (\ref{XRef-Equation-427221941}). First, the factor that
depends on the distance $\chi _{*}$ to the emitter can be written
in polar form as
\[
e^{i \left( 1-\mu \right) k \chi _{*}}-1=2 i e^{i( 1-\mu ) k \chi
_{*}/2}\sin ( \frac{\left( 1-\mu \right) }{2}k \chi _{*}) .
\]

Next, the polarization amplitudes $h_{+}^{0}$ and $h_{\times }^{0}$
will in general have different phases. Let $h_{s}^{0}=H_{s}e^{i
\alpha _{s}}$, where $H_{s}$ and $\alpha _{s}$ are respectively
the real amplitude and phase of polarization component $s$. Let
$\mathcal{K}\equiv \{k,\hat{k},H_{+},H_{-},\alpha _{+},\alpha _{-}\}$
denote the unknown state vector defining the tensor mode. $k$ is
the angular frequency, $\hat{k}$ the spatial propagation direction,
while $H_{\pm }$ and $\alpha _{\pm }$ are the amplitudes and phases,
respectively, of the two polarization components. Then, after some
algebra, we find that
\[
\sigma ^{+}( \hat{k},\hat{p}) h_{+}^{0}( k) +\sigma ^{\times }(
\hat{k},\hat{p}) h_{\times }^{0}( k) =T( \mathcal{K},\hat{p})  e^{i
\psi ( \mathcal{K},\hat{p}) },
\]

with amplitude
\begin{equation}
T( \mathcal{K},\hat{p}) \equiv {\left( {\left( \sigma _{+}H_{+}\right)
}^{2}+{\left( \sigma _{\times }H_{\times }\right) }^{2} +2\left(
\sigma _{+}H_{+}\right) \left(  \sigma _{\times }H_{\times } \right)
\cos ( \alpha _{+}-\alpha _{\times }) \right) }^{1/2}%
\label{XRef-Equation-430185825}
\end{equation}

and phase
\begin{equation}
\psi ( \mathcal{K},\hat{p}) \equiv \arctan ( \frac{\sigma _{+}H_{+}
\sin  \alpha _{+}+\sigma _{\times }H_{\times }\ \ \sin  \alpha _{\times
}}{\sigma _{+}H_{+} \cos  \alpha _{+}+\sigma _{\times }H_{\times
}\ \ \cos  \alpha _{\times }}) .%
\label{XRef-Equation-62873946}
\end{equation}

Finally, the polarization angle takes the sinusoidal form
\begin{equation}
\Omega _{\left( \operatorname{CE}\right) }^{\left( T\right) }( \tau
) =W_{\left( \operatorname{CE}\right) }( \mathcal{K},\hat{p},\chi
_{*}) \sin ( k \tau +\gamma _{\left( \operatorname{CE}\right) }(
\mathcal{K},\hat{p},\chi _{*}) ) ,%
\label{XRef-Equation-58224042}
\end{equation}

with amplitude
\begin{equation}
W_{\left( \operatorname{CE}\right) }( \mathcal{K},\hat{p},\chi _{*})
\equiv 2T( \mathcal{K},\hat{p}) \frac{\sin ( \frac{\left( 1-\mu
\right) }{2}k \chi _{*}) }{1-\mu }%
\label{XRef-Equation-51222731}
\end{equation}

and phase
\[
\gamma _{\left( \operatorname{CE}\right) }( \mathcal{K},\hat{p},\chi
_{*}) \equiv -\left( \frac{\left( 1-\mu \right) }{2} k \chi _{*}+\psi
( \mathcal{K},\hat{p}) \right) .
\]

The {\itshape polarization wiggling amplitude} $W_{(\mathrm{CE})}(
\mathcal{K},\hat{p},\chi _{*}) $ defines the polarization wiggling
response of polarized electromagnetic radiation emitted at distance
$\chi _{*}$ from the detector and in direction $\hat{p}$ to a plane
gravitational wave with state vector $\mathcal{K}$. The amplitude
$W_{(\mathrm{CE})}$ and phase $\gamma _{(\mathrm{CE})}$ vary with
the state $\mathcal{K}$ of the gravitational wave as well as the
position of the emitter relative to the detector. We see from eq.
(\ref{XRef-Equation-58224042}) that the polarization wiggling frequency
is equal to the gravitational wave frequency.

Let us take a look at the conditions under which the amplitude $W_{(\mathrm{CE})}$
vanishes. First of all, for inclinations $\mu =\pm 1$, the direction
of the emitted radiation is aligned with the direction of the gravitational
tensor mode; $\hat{p}=\pm \hat{k}$. From eqs. (\ref{XRef-Equation-42812371})
and (\ref{XRef-Equation-428123954}) follows that $\sigma ^{+}( \hat{k},\pm
\hat{k}) =\sigma ^{\times }( \hat{k},\pm \hat{k}) =0$, which implies
that
\[
W_{\left( \operatorname{CE}\right) }( \mathcal{K},\pm \hat{k},\chi
_{*}) =0.
\]

Hence, polarized radiation is insensitive to gravitational radiation
propagating in the same or opposite direction. 

Next, there are certain emitter distances that give vanishing $W_{(\mathrm{CE})}$
for $|\mu |<1$:
\begin{equation}
\chi _{*}= \frac{2\pi  N}{\left( 1-\mu \right) k}, N=1,2,...%
\label{XRef-Equation-62992125}
\end{equation}

Finally, from eq. (\ref{XRef-Equation-51222731}) follows that
\[
{\lim }_{k \chi _{*}\rightarrow 0}W_{\left( \operatorname{CE}\right)
}( \mathcal{K},\hat{p},\chi _{*}) =0.
\]

Thus, a detector set up to measure polarization wiggling of electromagnetic
radiation emitted by an emitter at distance $\chi _{*}$ will be
insensitive to gravitational radiation with wave length significantly
larger than the emitter distance $\chi _{*}$. 

With the exception of these cases as well as other border cases
where $T( \mathcal{K},\hat{p}) $ vanishes for certain combinations
of gravitational wave state $\mathcal{K}$ and polarized radiation
direction $\hat{p}$, $W_{(\mathrm{CE})}( \mathcal{K},\hat{p},\chi
_{*})  \neq 0$. Therefore, polarized electromagenetic radiation
is in general subject to polarization wiggling in the presence of
gravitational radiation.

\subsection{Wiggle Rate from a Tensor Mode - Distant Emitter}\label{XRef-Subsection-322204016}

Next, let us consider a distant emitter (DE), which we define as
an emitter of polarized electromagnetic radiation that is cosmologically
distant, with a scale factor at the time of emission that is significantly
less than 1. We will assume a conformally flat background with ${a(
\eta ) }^{\prime }\neq 0$ and evaluate the wiggle rate of eq. (\ref{XRef-Equation-4157300}).
In order to get analytical results and gain an understanding of
the results, we will make the simplifying assumption that the observed
electromagnetic radiation is emitted in a matter-dominated spacetime.
Although not a realistic model of our universe at later times, we
expect that this approximation is a reasonable approximation for
tensor modes of smaller scale as long as they are sufficiently damped
at the end of the matter-dominate era to not yield significant contributions
to the wiggling rate from later times. In addition, it is a useful
approximation to get a qualitative understanding of how gravitational
tensor modes induce polarization wiggling in an expanding universe.

It is known that a matter-dominated universe has a scale factor\ \ $a(
\eta ) ={(\eta /\eta _{0})}^{2}$, where $\eta _{0}$ is the time
at which $a( \eta _{0}) =1$. This enables us to evaluate the third
term on the right-hand side of eq. (\ref{XRef-Equation-4157300}).
Reparametrizing the integral in terms of the variable $x\equiv k
\eta $, we get
\[
i k \mu  \operatorname*{\int }\limits_{\eta _{*}}^{\eta }d\eta ^{\prime
}\frac{e^{-i k \left( 1-\mu \right)  \eta ^{\prime }}}{a( \eta ^{\prime
}) }=i k \mu  \eta ^{2}J( x,x_{*}) ,
\]

where
\[
J( x,x_{*}) \equiv \operatorname*{\int }\limits_{x_{*}}^{x}dt\frac{e^{-i
\left( 1-\mu \right) t}}{t^{2}}
\]

and $x_{*}\equiv k \eta _{*}$. The integral $J( x,x_{*}) $ can be
expressed in terms of the exponential integral $E_{2}$ as follows
\cite{Abramowitz-Stegun-1965}:
\[
J( x,x_{*}) =\frac{E_{2}( i \left( 1-\mu \right) x_{*}) }{x_{*}}-\frac{E_{2}(
i \left( 1-\mu \right) x) }{x}.
\]

We may use that $E_{n}( z) =e^{-z}/z+\mathcal{O}( z^{-2}) $ for
$|z|\gg 1$ \cite{Abramowitz-Stegun-1965}. The relationship between
the wiggle rates in the physical and conformal frames is given by
eq. (\ref{XRef-Equation-41415526}). For small-scale modes with $k
\eta _{*}\gg 1$, the wiggle rate in the physical frame can then
be derived from eq. (\ref{XRef-Equation-217162013}) and becomes
\begin{multline}
{\tilde{\omega }}_{\left( \operatorname{DE}\right) }^{\left( T\right)
}( \eta ) =\frac{\omega ^{\left( T\right) }( \eta ) }{a( \eta )
}\\
=\frac{1}{a( \eta ) }\mathrm{Re}\left\{  \frac{i k }{1-\mu }\left(
\frac{1}{a( \eta ) }-\frac{{e}^{i k( 1-\mu ) \chi _{*}}}{a( \eta
_{*}) }\right) e^{-i k \eta }\sum \limits_{s}\sigma ^{s}( \hat{k},\hat{p})
h_{s}^{0}( k) \right\} %
\label{XRef-Equation-322201919}
\end{multline}

to leading order in $1/k {\eta }_{*}$. It is evident from eq. (\ref{XRef-Equation-322201919})
that for radiation emitted early, while $a( \eta _{*}) \ll a( \eta
) =\mathcal{O}( 1) $, the ${a( \eta _{*}) }^{-1}$ terms on the right-hand
side of eq. (\ref{XRef-Equation-322201919}) will be dominant, and
the wiggle rate depends primarily on the tensor mode amplitude at
emission. This confirms our earlier claim that, for a distant emitter
in an expanding universe, the dominant contributions to the present
wiggle rate of polarized radiation emitted at early times come from
gravitational fields present at the time of emission. 

We will assume that the observation happens at present times with
$\eta \simeq \eta _{0}$ and $a( \eta ) =1$. To lowest order, proper
time $\tau $ of the observer can be expressed in terms of coordinate
time as $\tau =\eta -\eta _{0}+\mathcal{O}( h) $.\ \ 

Integrating to obtain the polarization angle $\Omega _{(\mathrm{DE})}^{(T)}$
in terms of proper time $\tau $ and assuming that $a( \eta _{*})
\ll a( \eta ) $ yields
\begin{equation}
\Omega _{\left( \operatorname{DE}\right) }^{\left( T\right) }( \tau
) =W_{\left( \operatorname{DE}\right) }( \mathcal{K},\hat{p},\chi
_{*}) \sin ( k \tau +\gamma _{\left( \operatorname{DE}\right) }(
\mathcal{K},\hat{p},\chi _{*}) ) ,%
\label{XRef-Equation-58224154}
\end{equation}

where
\begin{gather}
W_{\left( \operatorname{DE}\right) }( \mathcal{K},\hat{p},\chi _{*})
\equiv \frac{T( \mathcal{K},\hat{p}) }{a( \eta _{*}) \left( 1-\mu
\right) }%
\label{XRef-Equation-58224322}
\\\gamma _{\left( \operatorname{DE}\right) }( \mathcal{K},\hat{p},\chi
_{*}) \equiv -\left( \left( 1-\mu \right) k \chi _{*}+\psi ( \mathcal{K},\hat{p})
\right) ,
\end{gather}

and\ \ $T( \mathcal{K},\hat{p}) $ and $\psi ( \mathcal{K},\hat{p})
$ are given by eqs. (\ref{XRef-Equation-430185825}) and (\ref{XRef-Equation-62873946})
above.

We see from eq. (\ref{XRef-Equation-58224154}) that the polarization
wigging frequency is equal to the gravitational wave frequency.
This is similar to the result of eq. (\ref{XRef-Equation-58224042})
for a close emitter. The most striking difference between the results
of eqs. (\ref{XRef-Equation-58224042}) and (\ref{XRef-Equation-58224154})
is between the amplitudes $W_{(\mathrm{CE})}$ and $W_{(\mathrm{DE})}$
and their dependency on the emitter distance $\chi _{*}$. For a
close emitter, the wiggling amplitude oscillates with emitter distance,
and polarization wiggling is suppressed for emitter distances that
satisfy eq. (\ref{XRef-Equation-62992125}). This is not the case
for a distant emitter as the wiggling amplitude increases slowly
with emitter distance. The reason for this can be found by comparing
the close emitter wiggle rate $\omega _{(\mathrm{CE})}^{(T)}( \eta
) $ of eq. (\ref{XRef-Equation-62992410}) with the distant emitter
wiggle rate ${\tilde{\omega }}_{(\mathrm{DE})}^{(T)}( \eta ) $ of
eq. (\ref{XRef-Equation-322201919}). The factor $(1-e^{i (1-\mu
)k \chi _{*}})$ of eq. (\ref{XRef-Equation-58224042}) can be seen
as giving rise to two wiggling components with the same amplitude,
but with a phase difference of $(1-\mu )k \chi _{*}+\pi $. When
varying the emitter distance, these two components will interfere
destructively with a regular distance interval. Similarly, for a
distant emitter, the corresponding factor $({a( \eta ) }^{-1}-e^{i
(1-\mu )k \chi _{*}}{a( {\eta }_{*}) }^{-1})$ of eq. (\ref{XRef-Equation-322201919})
gives rise to two wigging components, but with very different amplitudes
and therefore no interference with varying emitter distance.

\subsection{Using Polarization Wiggling to Measure the State Parameters
of a Plane Gravitational Wave}

Given the results of this section, an obvious question is: Can electromagnetic
polarization be used to measure all parameters of the state vector
$\mathcal{K}$ of a plane gravitational wave? 

The state vector $\mathcal{K}\equiv \{k,\hat{k},H_{+},H_{-},\alpha
_{+},\alpha _{-}\}$ has 7 independent parameters. Since the frequency
of the polarization wiggling induced by a plane gravitational wave
is the same as the frequency of the gravitational wave itself, the
angular frequency $k$ of a gravitational wave can be measured by
simply measuring the frequency of the induced polarization wiggling.

The effect of the other 6 parameters are entangled in the amplitude
$W$ and phase $\gamma $. Assuming the polarized radiation is a plane
electromagnetic wave, both the wiggling amplitude $W$ and phase
$\gamma $ are measurable quantities. It should therefore, in principle,
be sufficient to determine the remaining 6 parameters of the state
vector $\mathcal{K}$ by measuring the wiggling amplitudes and phases
of three beams of polarized radiation emitted by emitters at known
positions from different directions.

\section{Conclusions}

{\itshape Polarization wiggling} is a direct gravitational effect
on the polarization of electromagnetic radiation. The effect is
defined in distinctly operational terms from the perspective of
a single observer: In the presence of a gravitational field along
the radiation trajectory, it manifests as a rotation of the polarization
axis of the observed radiation in the frame of observation between
subsequent observation events. Based on general results from an
earlier paper \cite{Tangen-2024-1}, we have in this paper studied
polarization wiggling induced by weak gravitational fields. Linear
gravity was used to analyze the polarization wiggling effect in
spacetimes with flat or conformally flat backgrounds. 

Spacetimes with conformally flat backgrounds is an extensive class
of spacetimes that includes most spacetimes of interest in cosmology.
They have one thing in common: their homogeneous backgrounds are
conformally related to 4-dimensional Minkowski space. This relationship\ \ gives
rise to\ \ a major simplification in the analysis of polarization
wiggling: We showed how it is possible to calculate the polarization
wiggle rates in the conformal frame and transform the results back
to the physical frame. The conformal frame is a much simpler geometry
than the physical frame due to its flat Minkowski background, and
the wiggle rates in the physical frame can be obtained by just multiplying
the conformal frame results with a scale factor. This simplification
arises when electromagnetic radiation is decomposed into plane waves.
Within the geometric optics limit, electromagnetic plane waves propagate
along null geodesics. Therefore, we can make use of the fact that
null geodesics are invariant under conformal transformations and
that Maxwell's equations in a 4-dimensional vacuum are also conformally
invariant. 

In linear gravity, the gravitational field, represented by the metric
perturbation $h_{\mu \nu }$, can be decomposed into scalar, vector
and tensor perturbations that are sourced and evolve independently.
We found that the polarization wiggle rates induced by scalar, vector
and tensor perturbations to the metric are gauge invariant quantities
that can be analyzed independently.

Scalar perturbations to the metric do not induce polarization wiggling,
so electromagnetic polarization is insensitive to the scalar degrees
of freedom of the gravitational field.

Vector perturbations to the metric, on the other hand, may induce
polarization wiggling. The vector-induced polarization axis wiggle
rate $\omega ^{(V)}$ is a direct measure of the longitudinal gravitomagnetic
field $\mathcal{B}$ and thereby the matter currents sourcing it.
The contribution to the polarization axis wiggle rate from vector
perturbations has two terms. The first term is the difference in
the angular frame dragging rate around the direction of propagation
between the frame of emission and the frame of observation. The
second term integrates the angular acceleration of the frame dragging
rate caused by non-stationary gravitomagnetic fields present along
the radiation trajectory. This implies that, for a stationary gravitomagnetic
field, the polarization axis wiggle rate simply measures the difference
in angular frame dragging rate between emission and observation.
Thus, stationary gravitomagnetic fields will not induce polarization
wiggling of any passing electromagnetic radiation if this radiation
was emitted and measured at locations with negligible gravitomagnetic
fields.

As an example of polarization wiggling induced by vector pertrubations,
we analyzed the polarization wiggling effect of a stationary axi-symmetric
gravitational field on polarized radiation emitted by an emitter
in orbit around the gravitational source. Assuming an emitter with
a known orbit around a gravitational source with unknown angular
momentum, it was shown that the angular momentum of the gravitational
source can be determined by measuring the polarization wiggle rate
emitted at three different positions along the orbit.

Tensor perturbations to the metric also induce polarization wiggling.
We studied the effect of a single gravitational plane wave (tensor
mode) with an arbitrary state of polarization. Two particular cases
were investigated in detail: A tensor mode propagating on a flat
background and in an expanding, matter-dominated universe. In both
cases, we showed how the state parameters of the tensor mode can
be determined by measurements of the polarization wiggling frequency,
amplitude and phase of electromagnetic radiation emitted by emitters
at known positions from at least three different directions. In
both cases, the polarization axis of the observed electromagnetic
radiation will wiggle with a frequency that equals the frequency
of the gravitational wave. Both amplitude and phase of the polarization
wiggling can be used to determine the properties of the tensor mode.
The amplitude and phase of the polarization wiggling vary with the
position of the emitter; the polarization wiggle rate is sensitive
to both direction of the electromagnetic radiation and distance
to the emitter. The wiggling amplitude vanishes for gravitational
waves with wave lengths much larger than the distance to the emitter.

A peculiar feature of the polarization wiggling from a tensor mode
in an expanding universe is that polarization wiggling captures
the properties of the tensor mode at the time of emission of the
electromagnetic radiation. While the tensor mode fades away in the
expanding universe, its properties at early times will be preserved
in the polarization wiggling of any passing polarized electromagnetic
radiation.

\acknowledgments{}

I would like to thank an anonymous reviewer for helpful comments,
in particular pointing out that the assumption that the rest frame
congruence is geodesic might not be needed. This resulted in important
simplifications and generalizations of the results of Section \ref{XRef-Section-4563844},
giving them much wider applicability.

\appendix

\section{Statements and Declarations}

{\bfseries Funding} \ \ \ No funding was received for this work.

{\bfseries Author Contribution} \ All work for this paper was done
by the author (K. Tangen). 

{\bfseries Conflict of Interest} \ \ The author declares that there
are no conflict of interests.

{\bfseries Data Availability Statement} No data were used or produced
in this work.

\end{document}